\documentclass[article]{paper}
\usepackage[total={6in, 8in}]{geometry}
\makeatletter
\providecommand{\maketitle}{}
\renewcommand{\maketitle}{%
	\par
	\begingroup
	\renewcommand{\thefootnote}{\fnsymbol{footnote}}
	\renewcommand{\@makefnmark}{\hbox to \z@{$^{\@thefnmark}$\hss}}
	\long\def\@makefntext##1{%
		\parindent 1em\noindent
		\hbox to 1.8em{\hss $\m@th ^{\@thefnmark}$}##1
	}
	\thispagestyle{empty}
	\@maketitle
	\@thanks
	\endgroup
	\let\maketitle\relax
	\let\thanks\relax
}

\providecommand{\@maketitle}{}
\renewcommand{\@maketitle}{%
	\vbox{%
		\hsize\textwidth
		\linewidth\hsize
		\vskip 0.1in
		\centering
		{\LARGE\bf \@title\par}
		\def\And{%
		\end{tabular}\hfil\linebreak[0]\hfil%
		\begin{tabular}[t]{c}\bf\rule{\z@}{24\p@}\ignorespaces%
		}
		\def\AND{%
		\end{tabular}\hfil\linebreak[4]\hfil%
		\begin{tabular}[t]{c}\bf\rule{\z@}{24\p@}\ignorespaces%
		}
		\begin{tabular}[t]{c}\bf\rule{\z@}{24\p@}\@author\end{tabular}%
		\vskip 0.3in \@minus 0.1in
	}
}
\makeatother

\usepackage[utf8]{inputenc} 
\usepackage[T1]{fontenc}
\usepackage[hidelinks]{hyperref}
\usepackage{amsmath}
\usepackage{amssymb}
\usepackage{mathtools}
\usepackage{acro}
\usepackage{booktabs}
\usepackage[table]{xcolor}
\usepackage{collcell}
\usepackage{hhline}
\usepackage{multirow}
\usepackage{pgf}
\usepackage{rotating}
\usepackage{tikz}
\usepackage{subcaption}
\usepackage{color}
\usepackage{lineno}
\usepackage{todonotes}

\usepackage{localmacros}

\DeclareAcronym{AR}{
  short = AR,
  long = Auto-regression}
\DeclareAcronym{ARIMA}{
  short = ARIMA,
  long = Auto-regressive Integrated Moving Average}
 \DeclareAcronym{SARIMA}{
  short = SARIMA,
  long = Seasonal Auto-Regressive Integrated Moving Average} 
 \DeclareAcronym{SARIMAX}{
  short = SARIMAX,
  long = Seasonal Auto-Regressive Integrated Moving Average with Exogenous Regresors} 
\DeclareAcronym{ES}{
  short = ES,
  long = Exponential Smoothing}
\DeclareAcronym{SVM}{
  short = SVM,
  long = Support Vector Machine}
  \DeclareAcronym{SVR}{
  short = SVR,
  long = Support Vector Regression}
\DeclareAcronym{NN}{
  short = NN,
  long = Neural Network}
\DeclareAcronym{RMSE}{
  short = RMSE,
  long = Root Mean Square Error}
\DeclareAcronym{ML}{
  short = ML,
  long = Machine Learning}
\DeclareAcronym{DL}{
  short = DL,
  long = deep learning}
\DeclareAcronym{LightGBM}{
  short = LightGBM,
  long = Light Gradient Boosting Machine}
\DeclareAcronym{AI}{
  short = AI,
  long = Artificial Intelligence}
\DeclareAcronym{XAI}{
  short = XAI,
  long = Explainable Artificial Intelligence}
\DeclareAcronym{ADF}{
  short = ADF,
  long = Augmented-Dickey Fuller}
\DeclareAcronym{ACF}{
  short = ACF,
  long = Autocorrelation Function}
\DeclareAcronym{PACF}{
  short = PACF,
  long = Partial Autocorrelation Function}

\begin{document}

\title{Neural Network Modeling for Forecasting Tourism Demand in Stopića Cave: A Serbian Cave Tourism Study}

\author{Buda Bajić \vspace{0.5mm} \\
  Faculty of Technical Sciences \\ 
  University of Novi Sad \\
  Serbia \\ 
  \href{mailto:buda.bajic@uns.ac.rs}{\texttt{buda.bajic@uns.ac.rs}}
  \And 
   Srđan Milićević \vspace{0.5mm} \\
  Faculty of Technical Sciences \\ 
  University of Novi Sad \\
  Serbia \\ 
  \href{mailto:srdjan88@uns.ac.rs}{\texttt{srdjan88@uns.ac.rs}} 
  \And 
  Aleksandar Antić \vspace{0.5mm} \\
  Institute of Geography and Sustainability \\ 
  University of Lausanne \\
  Switzerland \\ 
  \href{mailto:aleksandar.antic@unil.ch}{\texttt{aleksandar.antic@unil.ch}}
  \And 
  Slobodan Marković \vspace{0.5mm} \\
  Faculty of Sciences \\ 
  University of Novi Sad \\
  Serbia \\ 
  \href{mailto:slobodan.markovic@dgt.uns.ac.rs}{\texttt{slobodan.markovic@dgt.uns.ac.rs}} 
  \And
  Nemanja Tomi\'{c}\vspace{0.5mm} \\
  Faculty of Sciences \\ 
  University of Novi Sad \\
  Serbia \\ 
  \href{airtomic@gmail.com}{\texttt{airtomic@gmail.com}}}

\maketitle

\begin{abstract}

For modeling the number of visits in Stopića cave (Serbia) we consider the classical \ac{ARIMA} model, \ac{ML} method \ac{SVR}, and hybrid NeuralPropeth method which combines classical and \ac{ML} concepts. The most accurate predictions were obtained with NeuralPropeth which includes the seasonal component and growing trend of time-series. In addition, non-linearity is modeled by
shallow \ac{NN}, and Google Trend is incorporated as an exogenous variable. Modeling tourist demand represents great importance for management structures and decision-makers due to its applicability in establishing sustainable tourism utilization strategies in environmentally vulnerable destinations such as caves. The data provided insights into the tourist demand in Stopi\'{c}a cave and preliminary data for addressing the issues of carrying capacity within the most visited cave in Serbia.

\end{abstract}

\section{Introduction}

Modeling tourist demand includes a complex but necessary set of activities and analyses that can potentially determine market norms and directly shape tourist offers (\cite{a1_buhalis2005tourism}, \cite{a2_frechtling2012forecasting}). Research on tourism demand states that understanding tourist demand enables efficient allocation of resources, sustainability of revenue management, infrastructure planning, and risk management (\cite{a3_buhalis2000tourism}, \cite{a4_ritchie2003competitive}, \cite{a5_dwyer2009destination}, \cite{a6_vanhove2022economics}). Bearing in mind that this type of modeling indicates market trends, consumer behavior and preferences, management structures that manage tourist destinations can use this information to identify niche markets and new trends. Therefore, forecasting tourism demand can have significant impacts on maintaining optimal competitive markets within the tourism industry (\cite{a7_crouch1999tourism}, \cite{a2_frechtling2012forecasting}). Based on the prediction of demand dynamics, it is also possible to adapt competitive pricing strategies, within which prices at destinations can be increased and decreased depending on the expected tourist demand (\cite{a8_song2006tourism}, \cite{a9_martins2017empirical}, \cite{a10_li2019competitive}, \cite{a11_abrate2019impact}). In addition, \cite{a12_song2012tourism} argues that understanding tourist demand can influence the development of new products and services, which are compatible with the evolving needs and preferences of tourists. Tourist demand can also dictate the efficient use of marketing resources, in order to maximize reach and impact (\cite{a3_buhalis2000tourism}, \cite{a13_holloway2004marketing}, \cite{a14_sigala2012social}, \cite{a15_hudson2017marketing}), which is crucial for branding and competitiveness. Furthermore, operational efficiency is yet another factor on which tourist demand can have a significant impact. Mandal \cite{a16_mandal2018exploring} states that sustainable operational efficiency within the tourism industry largely depends on data-driven decision-making, thus exploring tourist demand is also a step towards enhanced productivity optimization. This includes managing inventory, schedules, and the number of employees (\cite{a17_lenny2007impact}, \cite{a18_lovelock2013strategies}, \cite{a19_shabanpour2018analysis}). The applicability of tourism demand modeling is especially evident when it comes to special forms of tourism affirmation (\cite{a20_burger2001practitioners}, \cite{a21_trauer2006conceptualizing}, \cite{a22_xie2021forecasting}). In the case of nature-based tourism (\cite{a23_dimitrov2013long}, \cite{a24_aliani2018modeling}, \cite{a25_rice2019forecasting}, \cite{a26_abu2021sarima}), forecasting tourist demand can be of great importance for adjusting carrying capacity measures in certain destinations. Numerous research (\cite{a27_o1986tourism}, \cite{a28_butler1999sustainable}, \cite{a29_mccool2001tourism}, \cite{a30_liu2003sustainable}, \cite{a31_fennell2004tourism}, \cite{a32_lobo2013projection}, \cite{a33_zelenka2014concept}, \cite{a34_lobo2015tourist}, \cite{a35_guo2017remaking}, \cite{a36_carrion2021environmental}, \cite{a37_cheablam2021assessment}, \cite{a38_sunkar2022geotourism}) indicates that carrying capacity is one of the most important indicators of sustainable and responsible tourism, especially when it comes to destinations that are highly vulnerable, both from natural processes and from anthropogenic influence. Therefore, predicting the increase in tourist demand can be of great importance for management structures, because it can indicate the need to implement certain measures to prevent overexploitation and over-tourism. 

In the last few decades, there has been a development of tourism of specialized interest, which focuses on geological attractiveness. Geotourism includes the affirmation of geologically significant landscapes and places that can have a certain market value obtained through the interpretation of knowledge (\cite{a39_gordon2018geoheritage}). Education and conservation of geodiversity are the primary elements of geotourism and as such have the most important role in the identification and valorization of geoheritage (\cite{a40_bentivenga2019geoheritage}). Therefore, geotourism through the transfer of knowledge provides value to geologically significant areas, both for the needs of tourism development (\cite{a41_dowling2006geotourism}, \cite{a42_chen2015principles}, \cite{a43_dowling2018geotourism}, \cite{a44_olafsdottir2019geotourism}) and for the effective implementation of geoconservation efforts (\cite{a45_brilha2002geoconservation}, \cite{a46_gray2005geodiversity}, \cite{a47_burek2008history}, \cite{a48_henriques2011geoconservation}, \cite{a49_crofts2020guidelines}, \cite{a50_williams2020geoconservation}).  

In the case of karst landscapes, which represent one of the most vulnerable areas in which tourist activities are carried out (\cite{a51_ruban2018karst}, \cite{a52_telbisz2020significance}, \cite{a53_zhang2023aesthetic}), geoconservation is a basic indicator of ethically-responsible use of karst resources \cite{a54_taheri2021human}. Within the karst areas, the sites that are mostly used for mass tourism are caves (tourist caves; i.e. show caves). A detailed study on global cave tourism that explored the number of tourist visits \cite{a55_chiarini2022global} indicates a very high number of visits to show caves. China boasts the highest annual visitation rate, with 19 million tourists to its cave destinations. In the United States, 9.9 million annual visitors have been recorded and within Europe, France stands out by having 5.2 million tourists annually to its caves, followed by Spain with 2.9 million visitors. Germany and Italy contribute significantly to the global cave tourism landscape, each hosting 2.4 million and 2.3 million tourists annually. Evidently, caves are a major focus of tourists around the world. Due to geoconservation standards and protection, it is necessary to pay special attention to modeling and monitoring the global tourist demand for cave tourism. Moreover, significant challenges within cave tourism are reflected primarily in the negative consequences that arise from the very arrangement of the cave for tourist use. This includes the installation of artificial lighting, construction, and introduction of substances harmful to the underground ecosystem \cite{a55_chiarini2022global}. In addition, the harmfulness of tourism for caves is reflected in the increase in subterranean temperature, CO2 levels, and changes in air humidity (\cite{a56_pulido1997human}, \cite{a57_baker1988environmental}, \cite{a58_sebela2015cave}, \cite{a59_novas2017real}, \cite{a60_constantin2021monitoring}). However, caves represent important destinations for multidisciplinary education, interpretation of human history, and environmental dynamics. For this reason, it is necessary to maximize the sustainable economic affirmation of caves, so that cave tourism is compatible with geoconservation standards. The advantage of management structures is that there are significant possibilities for monitoring and control within the caves themselves. In particular, visitors cannot walk outside the marked paths and cannot visit places in the cave that are not adequately lit and arranged for visiting without specialized equipment. Thus, monitoring is in most cases at a high level and this provides the possibility of effective quality control and the protection of the subterranean ecosystem. 

The aim of this paper is to model tourist demand for Stopića cave in West Serbia. In previous years, this cave had an exceptional increase in the number of tourist visits, and it became the most visited, surpassing the Resava cave, which for decades was the most visited in Serbia. This unique case represents an important local economic indicator that occurred as a result of the proximity of Zlatibor, which is a highly visited mountain center. The analysis includes a comprehensive time series dataset comprising the monthly visitation figures spanning from the year 2010 through 2023, thereby encompassing a total of 168 months of observational data. This temporal scope allows an exploration of visitor trends, facilitating forecasting methodologies to be employed effectively. Through modeling of these visitation patterns, we aim to gain insights that are essential for enhancing strategic planning and management practices in the context of Stopića cave's visitor economy. 

This paper is organized as follows. In section \ref{sec:related_work} we give an overview of different approaches used in the previous studies which aimed to forecast tourist arrivals. Section \ref{sec:method} presents models we use to forecast the number of visits to Stopića cave in Serbia in this study. Next, we present the experimental setup and results in section \ref{sec:eval_res}. Finally, the discussion is presented in section \ref{sec:discussion} and we conclude in section \ref{sec:conclusion}.

\section{Theoretical background}\label{sec:related_work}

Methods for forecasting time series can be divided into three categories: classical statistical methods, methods based on \ac{ML}, and hybrid methods which fuse both model and data-driven methodological approaches. 

The classical statistical forecasting methods were exhibiting the best performances before \ac{ML} methods started outperforming them, as demonstrated in several early time-series forecasting competitions, e.g., in M3 \cite{makridakis2000m3}. These methods attempt to identify patterns, trends, seasonality, and irregularities in the data observed over different time periods. They are particularly useful for understanding the underlying structure and pattern of the data and therefore offer interpretable forecasts for stakeholders. For forecasting tourism demand, the most widely used statistical forecasting method is \ac{ARIMA} and its versions which include seasonality and/or exogenous variables, see \cite{song2019review} and references therein. \ac{ES} is also used in many studies that forecast tourism demand (\cite{athanasopoulos2009hierarchical}, \cite{fildes2011evaluating}). 

In recent years, \ac{ML} techniques became popular for forecasting tourism demand, such \ac{NN} (\cite{claveria2015tourism}, \cite{chen2012forecasting}), \ac{SVR} (\cite{chen2007support},  \cite{chen2015forecasting}) and others. The most important advantage of data-driven methods is that they do not require stationarity or specific distribution of time series. Moreover, these models can explain non-linear relationships between input and output variables without a priori knowledge about them. However, the interpretability of these models is still an open research question. Also, in some applications, the amount of available data can be still too small for \ac{ML} techniques to train well so practitioners should carefully choose model complexity in order to avoid overfitting.

Hybrid methods bridge the gap between classical statistical and scalable \ac{DL} models by uniting them. Those methods are the best
performers in M4 forecasting competition \cite{makridakis2020m4}. In recent years, they are also used in many forecasting applications. For the purposes of tourism demand forecasting, in \cite{nor2018hybrid} \ac{ARIMA} and \ac{NN} are combined in order to forecast Malaysia's tourism demand. Similarly, \cite{abellana2021hybrid} combines \ac{SARIMA} and \ac{SVR} for modeling Philippine tourism demand. In this study, we consider modeling tourism demand in Stopića cave in Serbia by NeuralPropeth \cite{triebe2021neuralprophet} hybrid method. As baseline methods, we use \ac{ARIMA} as the most popular statistical/classical method and \ac{SVR} - frequently utilized \ac{ML} method for tourism demand forecasting.

Although the findings from the latest M5 time-series forecasting competition \cite{makridakis2022m5} demonstrate that modern pure \ac{ML} methods based on decision trees (such as \ac{LightGBM} method \cite{ke2017lightgbm}) now outperform
hybrid methods, in this study, due to the limited size of the time series, we do not consider such methods due to the risk of overfitting.

Apart from forecasting tourism demand exclusively based on its previous values, it is worth mentioning that many studies investigated how exogenous variables can help in predicting targeted time series. The most popular recently studied such explanatory variables are e.g. internet big data (e.g. Google Trend \cite{sun2019forecasting}, \cite{li2020forecasting}, \cite{gunter2016forecasting}, \cite{park2017short}, \cite{volchek2019forecasting}, \cite{clark2019bringing}) and social media and online reviews (e.g. TripAdvisor \cite{hu2022tourism}). We also consider Google Trends for modeling our time series.

\section{Method}\label{sec:method}

In the following subsections, a brief explanation of considered time series forecasting methods is given. 

\subsection{\acl{ARIMA}}\label{sec:arima}

Auto-Regressive Integrated Moving Average (\ac{ARIMA}) \cite{box2015time} model is one of the most frequently used models in time series analysis. The model is constructed to predict future trends of non-stationary data and represents an extension of the Auto-Regressive Moving Average (ARMA) model. It can be efficiently applied to eliminate trends and the non-stationarity of the mean using differencing between consecutive observations. 

\ac{ARIMA} model is generally denoted by \ac{ARIMA}$(p,d,q)$, where $p$ represents the order (number of lags) of the auto-regressive model, $d$ is the degree of differencing and $q$ denotes the order of the moving-average model. For given time series $\data_t,$ \ac{ARIMA}$(p,d,q)$ model is given by formula 
\begin{equation}\label{eq:ARIMA}
 \Big(1-\sum_{i=1}^{p}\arcoef_iL^i\Big)(1-L)^d \data_t=\Big(1+\sum_{i=1}^q \theta_i L^i\Big)\epsilon_t,   
\end{equation}
where $t$ is a positive integer, $L$ is the lag operator defined as $L^i\data_t=\data_{t-i},$ $\arcoef_i$ are the coefficients of the auto-regressive part of the model, $\theta_i$ are the coefficients of the moving average part and $\epsilon_t$ are error terms. The error terms $\epsilon_t$ are assumed to be independent with normal ${\cal N}(0,\sigma)$ distributions. There are several methods for determining values of parameters $p, d,$ and $q$ such as \ac{ADF} test, \ac{ACF}, and \ac{PACF}. 

As the number of tourist visits generally depends on the period of the year, for predictions of the number of tourists, it is useful to include the seasonal component in the model. Besides regular, seasonal data require seasonal differencing to become stationary. For this purpose, the \ac{SARIMA} model is used. \ac{SARIMA} model is denoted by \ac{ARIMA}$(p,d,q)(P,D,Q,M),$ where $M$ represents the seasonal period, i.e., number of observations per year, and $P, D$ and $Q$ are auto-regressive, differencing and moving average terms for the seasonal part of the model, respectively.

\ac{SARIMAX} model represents another generalization of \ac{ARIMA} model that includes both seasonality and exogenous variables. In this paper, the Google Trends data are used as one of the most popular tools in forecasting. The model has excellent performances which will be verified through results on tested data.

\subsection{\acl{SVR}}\label{sec:svr}

\ac{SVM} is \ac{ML} model %based on statistical learning frameworks 
initially developed for classification and later adjusted for regression (\ac{SVR}).
Here we briefly introduce \ac{SVR} \cite{SVR}, one of the most powerful techniques for solving both linear and nonlinear regression problems.

\begin{figure}[ht!]
\centering
\includegraphics[scale=0.2]{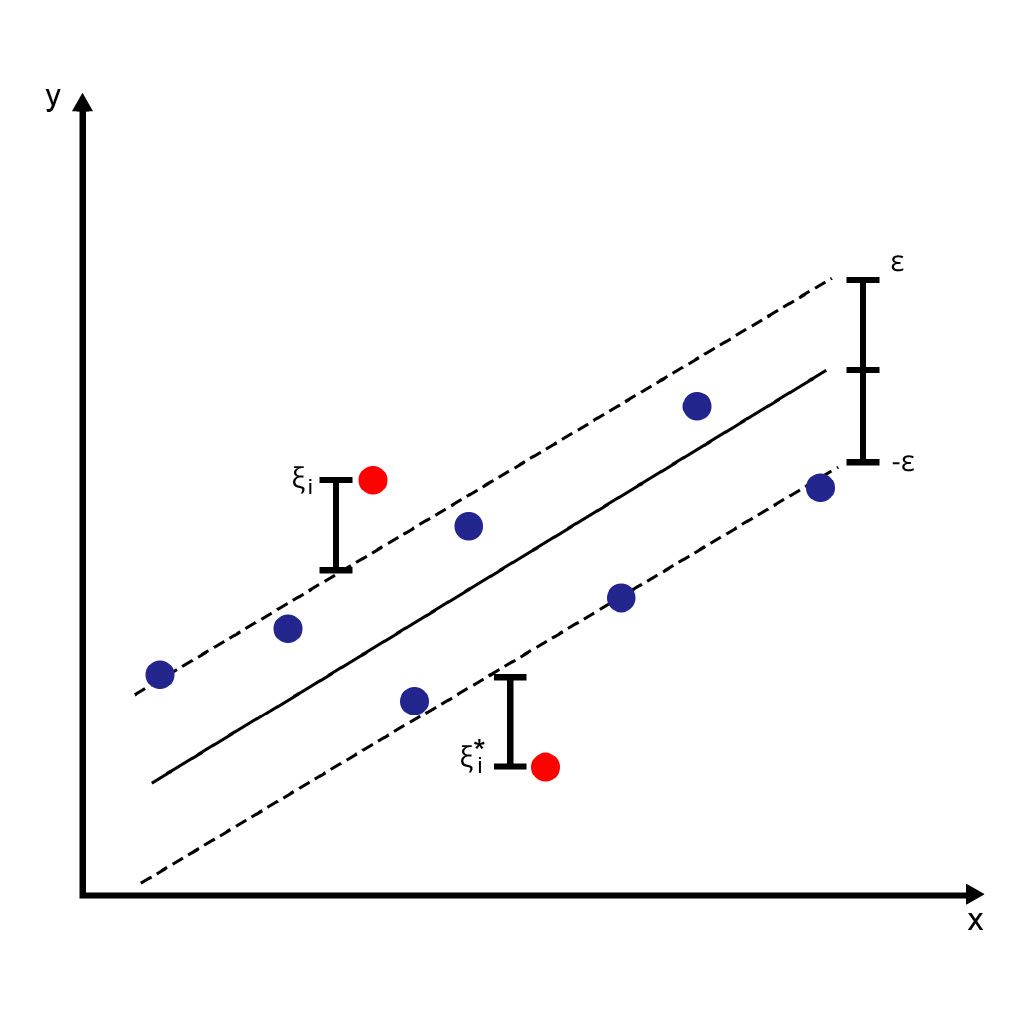}
\caption{\acl{SVR} model.}
\label{fig:svr_model}
\end{figure}

The linear regression model in general is given by
\begin{equation}\label{eq:regression}
%y = \alpha \cdot x + b,
y = \innerproduct{\alpha}{x}+ \beta,
\end{equation}
where $y \in \mathbb R$ is dependent variable, $x \in \mathbb R^m$ is independent variable, $\alpha \in \mathbb R^m$ and $\beta \in \mathbb R,$ are unknown coefficients and $\innerproduct{\alpha}{x}$ denotes the inner product between $\alpha$ and $x$. Classical linear regression models are based on estimating unknown coefficients for the given training set $D=\{(x_i, y_i)\}, \ i \in \{1,2,...,n\}$ by minimizing the sum of squared prediction errors (differences between the actual and the predicted values of the dependent variable).
\ac{SVR} model gives us the flexibility to define how much error is ``acceptable'' in finding prediction values. Instead of a simple regression line (or hyperplane in high dimensional spaces), the goal here is to find a tube (Fig. \ref{fig:svr_model}) on the distance (margin) $\epsilon$ from the line ($\epsilon$-insensitive tube). In that way, the model only cares about data outside the tube. In other words, the coefficients $\alpha$ and $\beta$, which in particular describe the relationships between $y$ and $x$, are found such that the prediction errors are minimized while the margin between the regression line and the closest data points is maximized at the same time. 

More concretely, the coefficients $\alpha$ and $\beta$ are in \ac{SVR} estimated by 
minimizing the regularized cost function under constraints:
\begin{align}\label{eq:SVR}
\text{minimize}\ \ &  \frac{1}{2}||\alpha||^2 + \gamma \sum_{i=1}^{n} (\xi_i + \xi^*_i) \nonumber
\\ & y_i-\innerproduct{\alpha}{x_i} - \beta \leq \epsilon + \xi_i \nonumber
\\ & \innerproduct{\alpha}{x_i} + \beta - y_i \leq \epsilon + \xi_i^* \nonumber
\\ & \xi_i, \xi_i^* \geq 0,\ i \in \{1,2,...,n\},
\end{align}
where $\gamma$ is the balancing parameter between the regularization term of the cost function and the training error calculated as the sum of $\xi_i$ and $\xi_i^*$, which are slack variables that represent positive and negative deviations outside $[-\epsilon,\epsilon]$ region (see Fig. \ref{fig:svr_model}). In order to solve the $(\ref{eq:SVR}),$ the dual quadratic problem is formed:
\begin{align}\label{eq:dual}
\text{maximize}\ \ &  \sum_{i=1}^{n} y_i(\lambda_i-\lambda^*_i) -\epsilon\sum_{i=1}^{n} (\lambda_i+\lambda^*_i)-\frac{1}{2}\sum_{i=1}^{n} \sum_{j=1}^{n} (\lambda_i-\lambda^*_i)(\lambda_j-\lambda^*_j)\innerproduct{x_i}{x_j} \nonumber
\\ & \sum_{i=1}^{n} (\lambda_i-\lambda^*_i)=0\nonumber
\\ & 0 \leq \lambda_i,\lambda_i^* \leq \gamma,\ i \in \{1,2,...,n\},
\end{align}
where $\lambda_i$ and $\lambda_i^*$ are Lagrange multipliers that satisfy $\lambda_i\lambda_i^*=0.$
Finally, the decision function $(\ref{eq:regression})$ has the following explicit form:
\begin{equation}\label{eq:SVReq_nonlin}
y=\sum_{i=1}^{n} (\lambda_i-\lambda_i^*)K(x,x_i) + \beta,
\end{equation}
where $K(x,x_i) = \innerproduct{x}{x_i}$ in the linear case. In the non-linear case, $K$ represents the kernel function that transforms the data in a higher dimensional space to be suitable for linear separation, e.g., polynomial kernel ($K(x,x_i)=\innerproduct{x}{x_i}^d$) and Gaussian ($\displaystyle K(x,x_i)=e^{-\frac{||x-x_i||^2}{2\sigma^2}}$).

When \ac{SVR} is applied for time series forecasting, the independent variable $x$ contains time series lags, and the dependent variable $y$ is the next observation in time series.

\subsection{NeuralPropeth}\label{sec:neural_propeth}

In this study we deploy NeuralProphet \cite{triebe2021neuralprophet} as a hybrid time series forecasting method. It is an extension of Facebook's Prophet \cite{taylor2018forecasting}, it provides information for interpreting outputs (predictions) from internal parts of the model and therefore it belongs to \ac{XAI} methods. Interpretability of NeuralProphet is achieved thanks to the fact the model is based on an additive decomposition of time series. It combines the classic time series components with scalable \ac{NN} blocks and in that way it is able to fit non-linear relationships. Two such \ac{NN} modules are the auto-regression and covariate components and thanks to them it demonstrates better predicting accuracy in comparison to Facebook's Prophet.

More formally, the NeuralProphet decomposes the time series in multiple additive components where each produces $h$ future predictions at the same time.
For a single time step forecast ($h = 1$), the model is given as:
\begin{equation} \label{eq:neural_propeth}
\dataforecast_t = \trend(t) + \season(t) + \autoreg(t) + \exogenouspast(t) + \exogenousfuture(t) + \event(t)
\end{equation}
where $\trend(t)$ is the trend at time $t$, $\season(t)$ models the seasonal effects at time $t$, $\autoreg(t)$ includes the auto-regression effects at time $t$
based on past observations of the time series of interest, $\exogenouspast(t)$ captures the regression effects at time $t$ for lagged observations of exogenous variables (covariates), $\exogenousfuture(t)$ accounts for the regression effect of
future-known exogenous variables at time $t$ and $\event(t)$ represents effect
of certain events and holidays at time $t$. Each of the described components can be excluded if it is not relevant to the targeted time series.

\textit{The trend} is modeled in a classic way, as a piece-wise linear function with the growth rate which can change at predefined number of points, so-called changepoints (model hyperparameter). 

\textit{Seasonal component} is modelled by Fourier terms \cite{harvey199310}, with $m$ terms for seasonality with
periodicity $l$:
\begin{equation} \label{eq:seasonality}
\season_l(t) = \sum_{j=1}^m \left( a_j \cos \left(\frac{2 \pi j t}{l}\right) + b_j \sin \left(\frac{2 \pi j t}{l}\right) \right).
\end{equation}
Number of Fourier terms is by default set to be $m = 6$ with $l = 365.25$ for yearly seasonality, $m = 3$ with $l = 7$ for weekly seasonality,
and $m = 6$ with $l = 1$ for daily seasonality. Mode details can be found in \cite{triebe2021neuralprophet}.

\textit{\ac{AR}} predicts the future values of the target variable by using a linear combination of its past values. The auto-regressive model \ac{AR}($p$) is defined as:
\begin{equation}\label{eq:ar}
\displaystyle \data_t = s + \sum_{i=1}^p \arcoef_{i}\data_{t-i} + \datanoise_t, 
\end{equation}
where $p$ is the number of linearly combined past time steps and intercept is denoted with $s$. Coefficients $\arcoef_i$ control the direction and power/significance of included past values on the future value and $\datanoise_t$ is the noise term. Classical \ac{AR} model produces only one prediction ($h = 1$). Therefore, for prediction horizon with number of steps $h > 1$, $h$ classical \ac{AR} models have to be estimated. The \ac{AR} module in NeuralProphet is based on a modification of \ac{AR}-Net \cite{triebe2019ar}, which allows single model to make $h$ forecast steps for $h > 1$. Three types of auto-regression - linear, deep and sparse, can be considered within \ac{AR} module. Linear \ac{AR} is single \ac{NN} with only one layer which has $p$ inputs, $h$ outputs, and it does not have biases nor activation functions, so it is essentially same as classic statistical \ac{AR}. Deep \ac{AR} consists of a fully connected \ac{NN} with arbitrary number of hidden layers and non-linear activation functions (such as rectified linear unit (ReLU)) after each layer apart from the final one. The first layer inputs are $p$ last observations, the outputs of the final layer are $h$ future values, whereas number of hidden layers and number of neurons in them is controlled by the user. Finally, sparse \ac{AR} allows \ac{AR} order $p$ to be chosen as higher at the beginning, and then with use of a regularization only a few past observations can be forced to have weights which are not equal to $0$. It is merely a way of selecting the most significant time series lags.

\textit{The lagged regressor} component is almost same as the \ac{AR} component - 
the only difference is that the inputs are the past values of exogenous variable instead of the targeted time series. An individual lagged regressor component has to be made for each covariate if there are multiple.

\textit{Future regressors} component is same as the lagged regressor, except that we need to know the future values of exogenous variable and not only its past values. 

Two types of \textit{events and holidays} can be considered: user-defined events, where the user feeds the model with information about an uncommon events, or country-specific holidays, where the user only provides the name of a country and the model automatically takes into account its national holidays. In both scenarios, events and holidays are binary variables with values $1$ when the event occurs and $0$ otherwise. 

In case \ac{NN} modules are deployed within NeuralPropeth, the Huber loss function during training is optimized by PyTorch optimizers where the user can define all relevant training hyperparameters such as learning rate, number of epochs, batch size, etc. 

\section{Evaluation and Results}\label{sec:eval_res}

\subsection{Data description and experiment design}

\begin{figure}[ht!]
\centering
\includegraphics[width=1\linewidth]{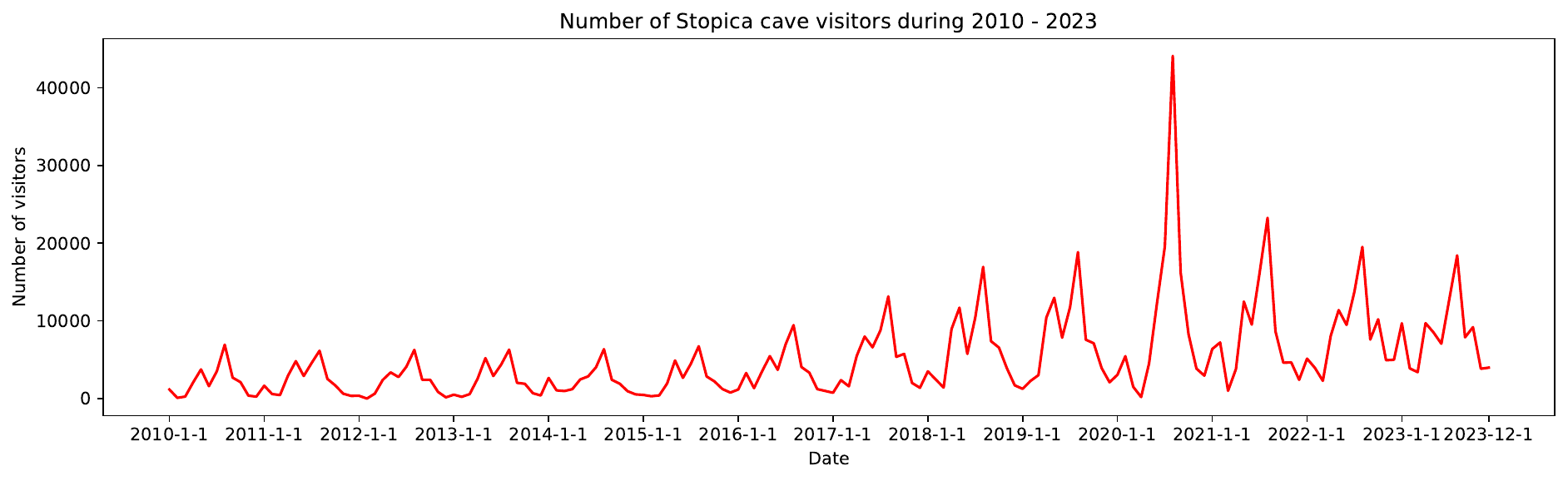}
\caption{Number of Stopića cave visitors during 2010-2023, monthly frequency.}
\label{fig:dataset}
\end{figure}

Forecast modeling of tourism demand for Stopića cave included the use of time series with the number of visitors for each month during 2010 - 2023 (168 months in total). Fig. \ref{fig:dataset} shows how the number of visitors changed during the entire period. It can be observed that this time series has a strong seasonal component, meaning that during the summer period (July and August) when many people go on summer vacations, yearly peaks occur, whereas during winter time the number of visits is much lower in comparison to summertime. This pattern is visible during the entire period and it repeats each year. Apart from the seasonal component, that growing trend is also, especially in the second half of the considered time frame. Another interesting event to notice is the highest number of visits that happened during the summer of 2020. It was during the COVID-19 pandemic, and Serbia was in partial lockdown in that period. Many countries had traveling restrictions during that year, so vacations were spent mainly in the country of origin. This happened to domestic tourists in Serbia, they were not able to travel abroad easily during that year so they spent holidays in Serbia and it influenced that the highest number of visits in history of Stopića cave occurred at that time.

\begin{figure}[ht!]
\centering
\includegraphics[width=1\linewidth]{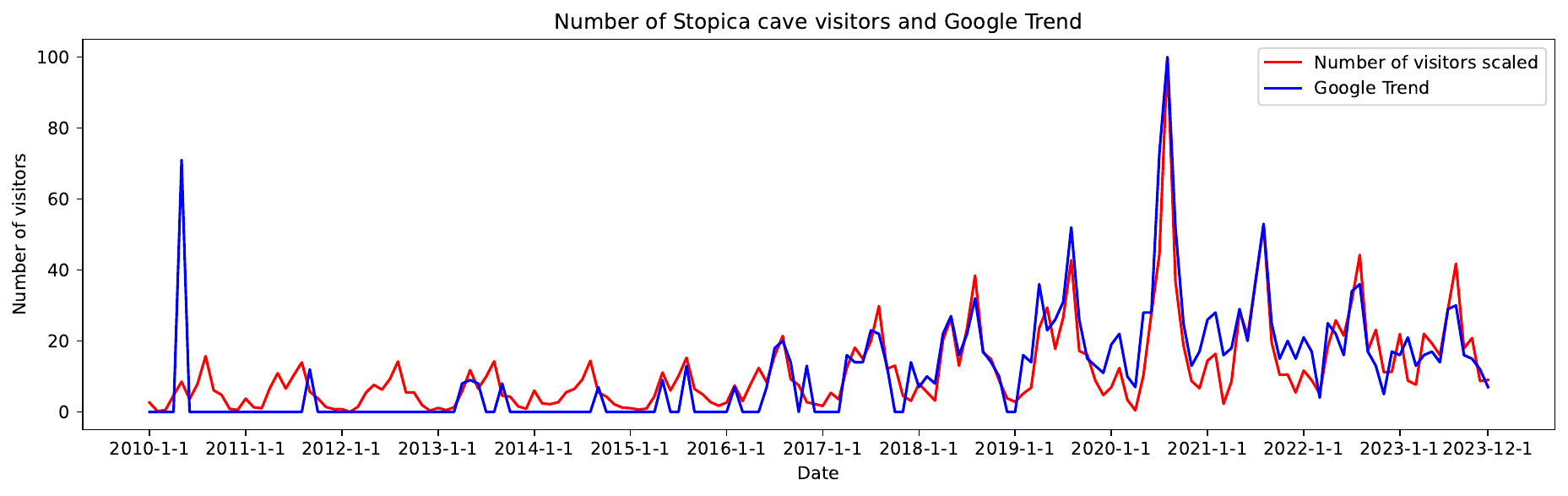}
\caption{Number of Stopića cave visitors during 2010-2023 (scaled to $[0,100]$) and Google Trend index.}
\label{fig:google_trend}
\end{figure}

Apart from an official number of visitors, we downloaded the Google Trend\footnote{https://trends.google.com/} index for the keyword ``Stopića pećina'' (pećina meaning cave in Serbian) for the considered period. Since most of the tourists who visit Stopića cave are domestic tourists (more than 95 \%), we opted for the keyword Serbian name of the cave. Obtained time series has also monthly frequency, and it measures the search volume of the chosen keyword. The search volume index exhibits search interest. It has values from $[0, 100]$ where value of $100$ corresponds to highest popularity of the keyword. Fig. \ref{fig:google_trend} shows the Google Trend index together with a number of visits scaled to the same range for better visibility. As can be observed from Fig. \ref{fig:google_trend}, it seems that the two time series are strongly correlated having similar seasonality, trend, and peaks. It seems that many visitors to the cave were searching the name of the cave on the web, either slightly before their planned visit or at the same time. In considered models, we try to include this series as an exogenous variable.

\subsection{Evaluation}

For evaluation of considered methods, we split the time series with the number of visitors into two parts - the first 156 months (period 2010 - 2022) were used for training \ac{ML} methods and the rest 12 months (year 2023) were used for testing all methods. For the testing phase, we predict/forecast the number of visitors and compare predicted values with actual ones.

For comparison of forecasted number of visitors with real ones, we choose \ac{RMSE} defined as:
\begin{equation}
\text{RMSE} = \sqrt{\frac{1}{T}\sum_{t=1}^T \left(\data_t - \dataforecast_t\right)^2} 
\end{equation}
where $T$ is the size of the data used in evaluation ($T = 12$ in our case), and $\data_t$ and $\dataforecast_t$ are actual and predicted number of visitors at time $t$. The smaller the measure is, the closer real and predicted values are. When comparing predictions of different models, the model with the smallest \ac{RMSE} is considered the best.

\subsection{Results}

\begin{figure*}[t!]
    \centering
    \begin{subfigure}[t]{1\textwidth}
        \centering
        \includegraphics[scale=0.45]{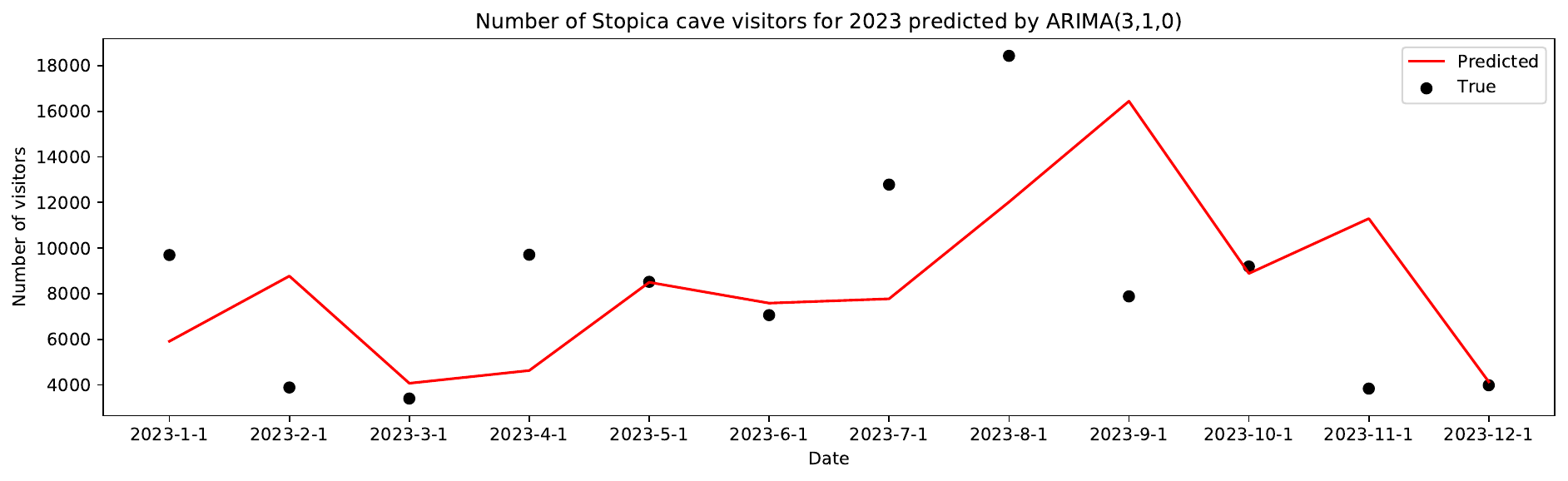}
        \caption{$\text{RMSE} = 4652.32$}
    \end{subfigure} \\
    \begin{subfigure}[t]{1\textwidth}
        \centering
        \includegraphics[scale=0.45]{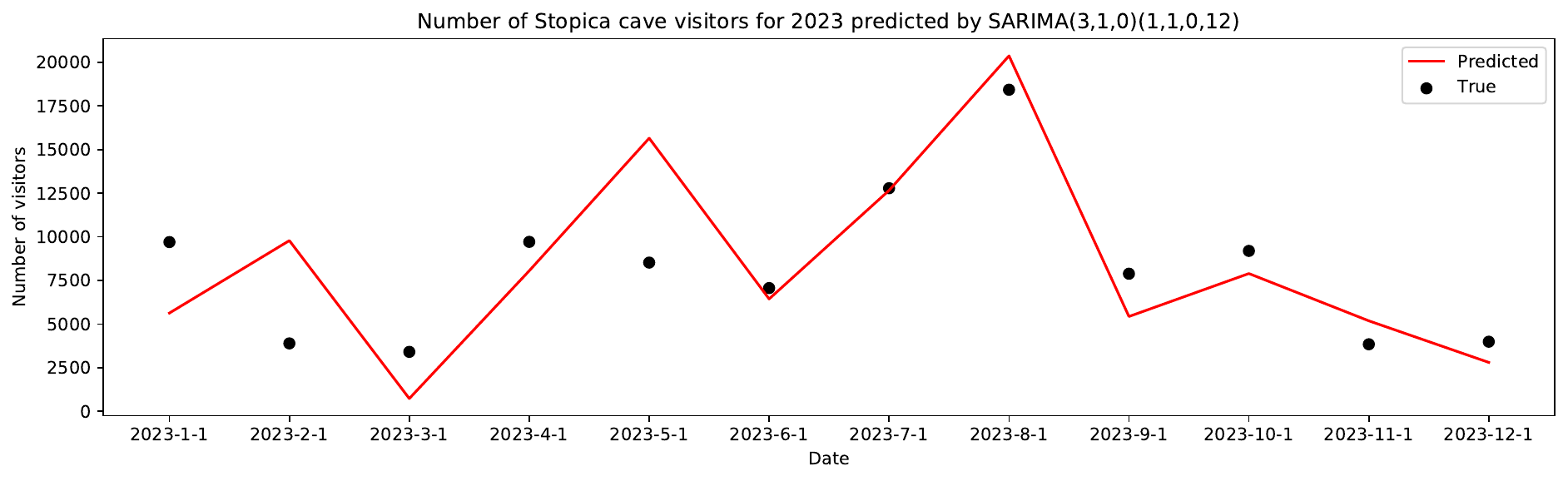}
        \caption{$\text{RMSE} = 3254.70$}
    \end{subfigure} \\
    \begin{subfigure}[t]{1\textwidth}
        \centering
        \includegraphics[scale=0.45]{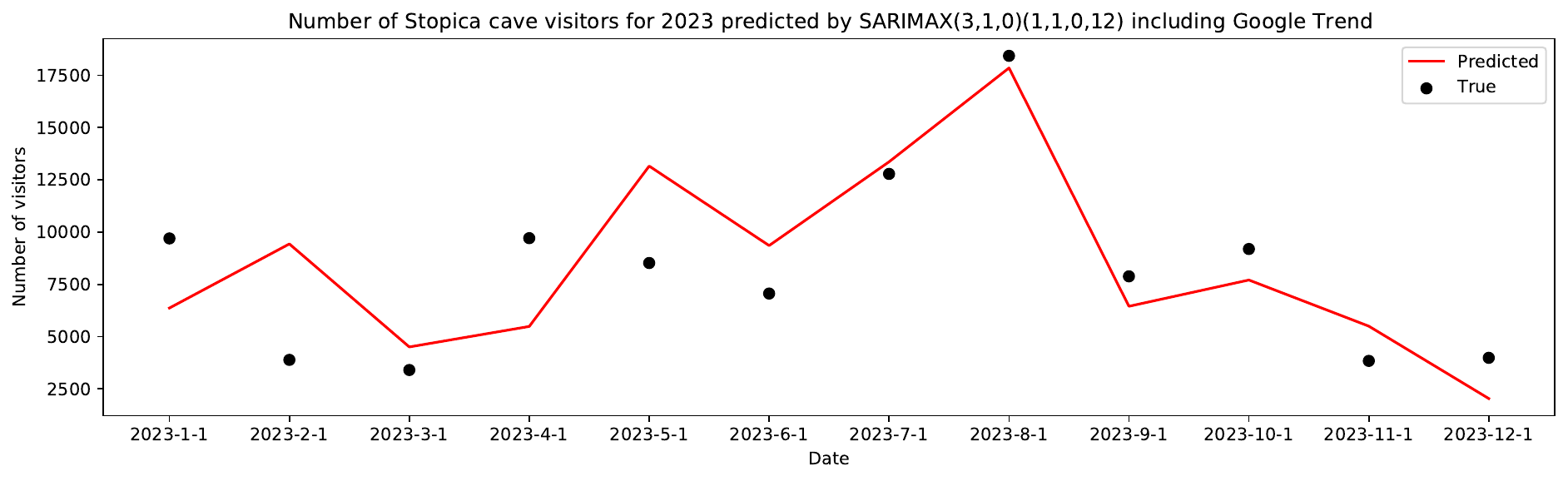}
        \caption{$\text{RMSE} = 2873.99$}
    \end{subfigure}
    \caption{Prediction of number of visitors for 12 months for 2023 obtained with different versions of \ac{ARIMA} model. The most accurate predictions gave SARIMAX - \ac{ARIMA} which includes a seasonal component and Google Trend as an exogenous variable.}
    \label{fig:arima}
\end{figure*}

The first model we consider is \ac{ARIMA}. Inspecting \ac{ACF} and \ac{PACF}, we concluded that the number of lags (order of auto-regressive model) that should be included in the model equals $p=3$, whereas the degree of differencing should be $d=1$ and order of moving-average $q=0$. In order to predict the entire 12 months, we fit 12 \ac{ARIMA}$(3,1,0)$ models since a single model can only predict the number of visitors for one month ahead. The plot of the actual vs. predicted number of visitors for 12 months during 2023 is given in Fig. \ref{fig:arima} (a), and comparing predictions with true data gave $\text{RMSE} = 4652.32$. Further, we included in the same model also seasonal component for which we use single lag $P=1$, degree of differencing $D=1$, order of moving-average $Q=0$, and $M=12$ since we have monthly data. Fig. \ref{fig:arima} (b) shows that including seasonal component into \ac{ARIMA} gave more accurate predictions as \ac{RMSE} significantly decreased to $\text{RMSE} = 3254.70$. Finally, we included Google Trend as an external regressor and this led to a further decrease of $\text{RMSE} = 2873.99$ and gave the best fit among all considered \ac{ARIMA} variants.

\begin{figure}[ht!]
\centering
\includegraphics[scale=0.45]{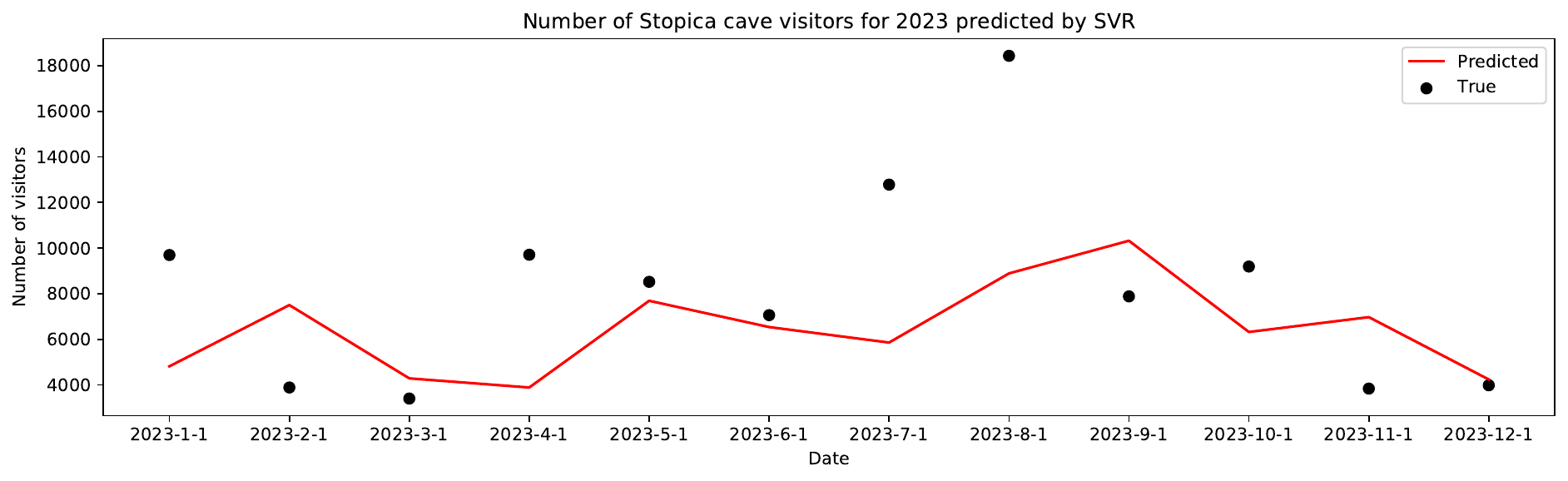}
\caption{Prediction of number of visitors for 12 months for 2023 obtained with \ac{SVR} model, $\text{RMSE} = 4430.33$.}
\label{fig:svr_res}
\end{figure}

Next, we train \ac{SVR} on monthly data from 2010-2022. As already mentioned at the end of Sec. \ref{sec:svr}, for independent variable $x$ we use time series lags. To make a fair comparison between different models, here we also consider $3$ past values of time series to be used for predicting future ones. \ac{SVR} with Radial Basis Function (RBF) kernel, regularization parameter $C=10$ and $\epsilon = 0.05$ tube is fitted, and predictions obtained with this model are presented in Fig. \ref{fig:svr_res}. Computed \ac{RMSE} $4430.33$ is a little bit better than \ac{RMSE} obtained with pure \ac{ARIMA}, but it is worse than estimated \ac{SARIMA} and \ac{SARIMAX} models. 

\begin{figure}[ht!]
\centering
\includegraphics[scale=0.45]{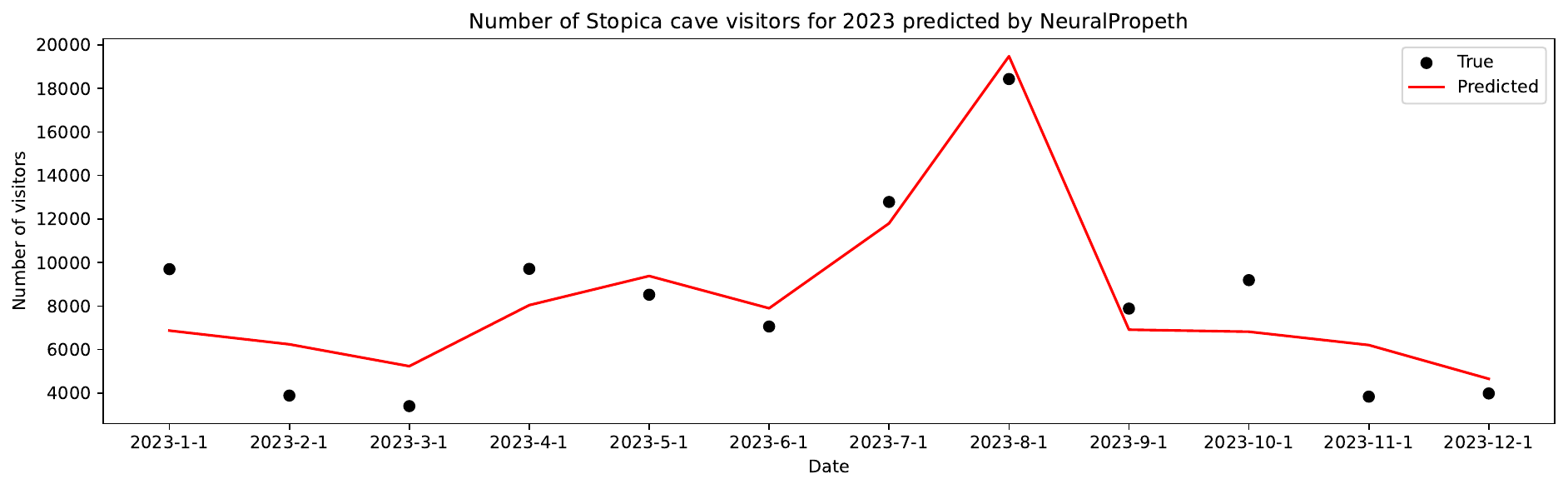}
\caption{Prediction of number of visitors for 12 months for 2023 obtained with the NeuralPropeth model, $\text{RMSE} = 1726.88$.}
\label{fig:neural_propeth}
\end{figure}

Finally, we experimented with the NeuralPropeth model. We included in the model yearly seasonality, a growing trend with the default number of trend changepoints, $3$ lags of targeted time series, and $2$ lags of Google Trend as an external lagged regressor. For modeling non-linearity, \ac{NN} with 2 hidden layers containing $4$ and $2$ nodes, respectively, is included in the model. Here we intentionally choose \ac{NN} of small size in order to prevent overfitting since the data we have has a relatively small number of observations from \ac{ML} perspective. The model is optimized with PyTorch AdamW optimizer with a learning rate of $0.003$. Fig. \ref{fig:neural_propeth} presents an actual and predicted number of visitors for 2023 obtained with the NeuralPropeth model. As we can see from the plot, and also by comparing NeuralPropeth \ac{RMSE} with \ac{RMSE} of previous models, the best fit is obtained by the estimated hybrid NeuralPropeth model with the chosen parameters explained above. Computed \ac{RMSE} for this model equals to $\text{RMSE} = 1726.88$ and it is approximately $40\%$ lower than the smallest \ac{ARIMA} models \ac{RMSE} (the one which \ac{SARIMAX} gave) or more than $60\%$ lower than \ac{RMSE} obtained by \ac{SVR}. 
 
\begin{figure}[ht!]
\centering
\includegraphics[scale=0.45]{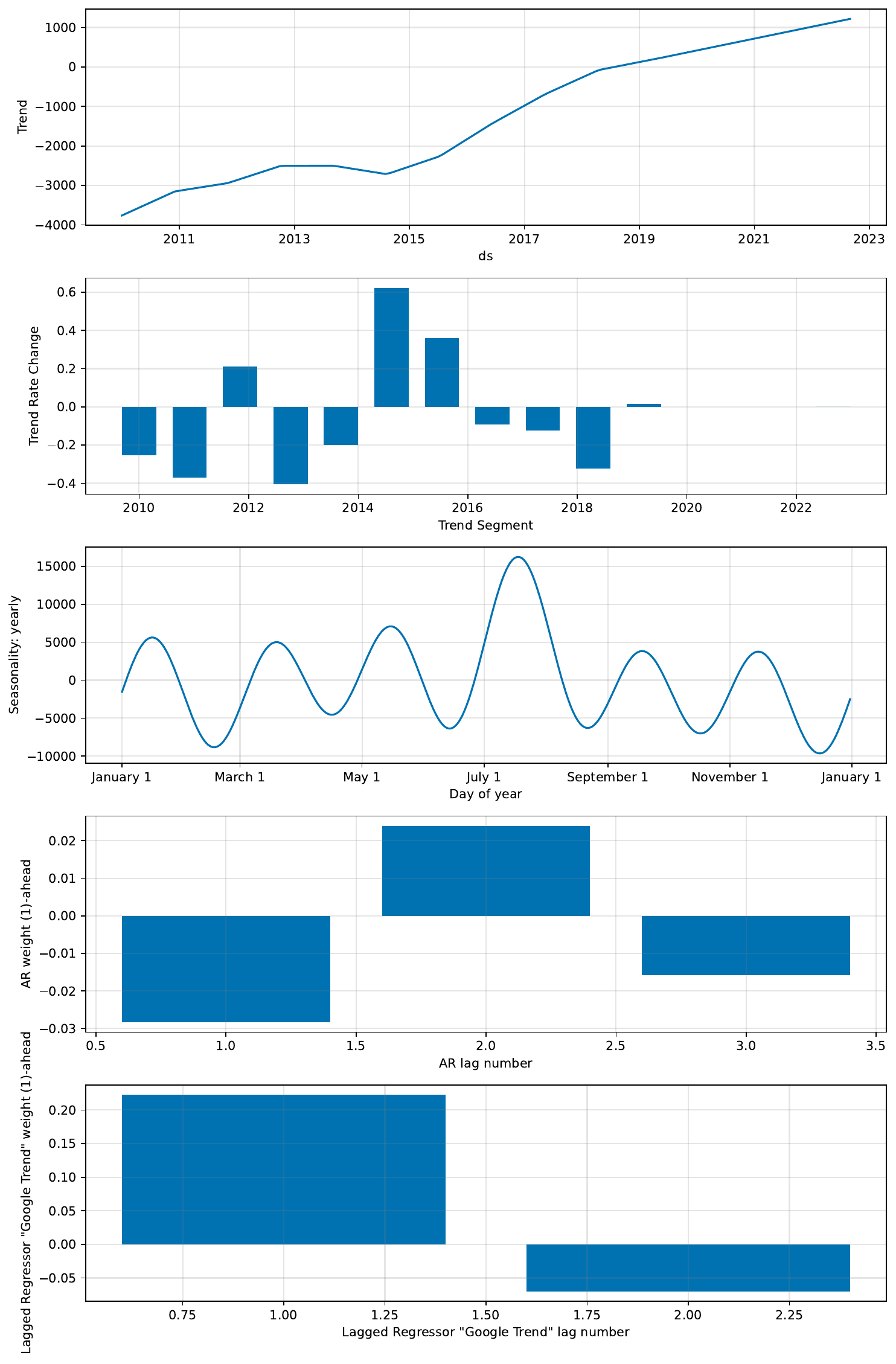}
\caption{Estimated parameters of NeuralPropeth model.}
\label{fig:neural_propeth_params}
\end{figure}

Fig. \ref{fig:neural_propeth_params} shows estimated trend and seasonal components as well as parameters for included $3$ and $2$ lags of targeted time series and Google Trend, respectively. The possibility to extract estimated model parameters is of great importance for stakeholders and policymakers, and it is an additional advantage of the NeuralPropeth model since many \ac{ML} based models are of a ``black-box'' nature for experts from the field of interest.    

\section{Discussion}\label{sec:discussion}

\subsection{Insights into Stopića cave tourism demand}\label{sec:sub_con1}

Cave tourism includes unique opportunities and challenges that require specialized strategies for effective management. The conducted analysis of the touristic demand of Stopića cave indicates the dynamism of the demand for the most visited cave in Serbia, and gives touristic implications that may be of importance to tourist organizations and decision-makers. Similar to many other destinations, Stopića cave also has visitation patterns that are influenced by seasonality, external events, and visitor preferences. The observed peak periods of visits during the summer represent the importance of adapting to seasonal growth, which includes optimizing the visitor experience. The increase in visits during the summer months of 2020 is the result of the COVID-19 pandemic, which is associated with travel restrictions, and the emphasis on the need for adaptive management for domestic tourism. The most significant increase in visits to the Stopića cave is the proximity of the mountain/tourist center Zlatibor. During the pandemic, many visitors stayed at this tourist center, which offers tourist activities throughout the season. A visit to the Stopića cave is one of the optional trips from the Zlatibor tourist center, which are often carried out as individual trips or as part of organized group excursions offered by tourist agencies. This increase in the number of visits directly affects the sustainability of Stopića cave as a tourist-accessible cave.

The sustainability of cave tourism requires a delicate balance between visitor access and conservation. As increased tourism demand can complicate carrying capacities at destinations, the results of our analysis serve as crucial preliminary inputs for the initial assessment of Stopića cave's carrying capacity. By quantifying tourism demand patterns, seasonal variations, and visitation trends, we gain valuable insights that can inform actionable steps toward carrying capacity estimates and the development of effective management strategies. However, it is essential to translate these insights into specific, applicable actions to ensure the accuracy and reliability of carrying capacity determinations and promote sustainable cave tourism management. For this, it is necessary to conduct ecological surveys and impact assessments. Nevertheless, the data shows future peak visitation levels, thus periods of potential overcrowding. This information is crucial for managing visitor access, optimizing tour routes, and implementing crowd control measures to prevent ecological degradation in sensitive areas. Recognizing these high-impact periods, tourism authorities can implement proactive management measures, such as visitor quotas, timed entry tickets, or temporary closures, to prevent environmental degradation and preserve the integrity of the subterranean ecosystem.

Furthermore, patterns in online search activity indicate a strong connection with the actual number of visitors, which further shows evident public curiosity and potential visitation intent. Exploring online search behavior can guide targeted marketing strategies and promotional efforts aimed at attracting visitors to the cave. Therefore, by gaining insights into online search, cave management can anticipate visitor trends and thus generate sustainable adaptive management strategies.

\subsection{Tourism demand and cave monitoring}\label{sec:sub_con2}

An insight into the tourist demand for the cave enables the planning of adequate measures of environmental monitoring in order to determine the dynamics of anthropogenic influence. With the increase in demand, it is necessary to establish continuous monitoring of climatic parameters that may occur due to the increased presence of visitors. Measuring changes in temperature, air humidity, and CO2 emissions are the most important factors that indicate that the increased visitation of the cave affects its ecosystem.

Increased tourist demand also indicates the importance of security measures. Thus, it is necessary to increase the safety of visitors through the implementation of adequate infrastructure, the functionality of which should be evaluated frequently. Safety also refers to the protection of the cave itself. Increased demand reflects a greater number of visitors, therefore it is necessary to introduce precautionary measures in order to maximize the protection of fragile aspects of the cave such as speleothems and groundwater quality.

\section{Conclusion}\label{sec:conclusion}

Three different methods are explored for modeling tourist arrivals in Stopića cave in Serbia on a monthly basis - classical \ac{ARIMA} with and without seasonal component and Google Trend as exogenous variable, pure \ac{ML} method \ac{SVR} and hybrid NeuralPropeth method which combines classical and \ac{ML} concepts. The best fit for the chosen test period of one year is obtained with NeuralPropeth which includes the seasonal component, growing trend, non-linearity modeled by shallow \ac{NN}, and Google Trend as an exogenous variable. The estimated NeuralPropeth model apart from giving the best predictions for considered time series, it outputs also the significance of the influence of lags for both auto-regressive and exogenous variable parts, helping policymakers in that way to better understand the model and consider using it further while taking important decisions.

The obtained research results have important implications for the touristic affirmation of caves. The implementation of advanced forecasting modeling enables management structures to make various strategic moves such as sustainable management of resources and conservation efforts. The use of such analytical techniques indicates effective modeling approaches that are crucial for the sustainability and protection of subterranean karst environments. Observed trends in tourism demand and the impact of factors such as seasonality and external events point to the fact that a continuous increase in tourist visitation to Stopića cave is to be expected. Due to this prediction, it is necessary to establish adequate protection measures that can ensure long-term subterranean environmental sustainability. This primarily involves monitoring microclimate indicators such as temperature fluctuations, air humidity, and CO2 emissions. The analysis of monitoring results can greatly contribute to the understanding of the anthropogenic impact on Stopića cave, as well as the level of its vulnerability. Establishing monitoring programs and tracking visitor trends are crucial for tourism authorities so they can assess the effectiveness of carrying capacity measures and adapt management strategies in order to ensure the long-term sustainability of Stopića cave as a tourist destination.

\section*{Acknowledgment}

This research has been supported by the Ministry of Science, Technological Development and Innovation (Contract No. 451-03-65/2024-03/200156) and the Faculty of Technical Sciences, University of Novi Sad through the project ``Scientific and Artistic Research Work of Researchers in Teaching and Associate Positions at the Faculty of Technical Sciences, University of Novi Sad'' (No. 01-3394/1). This research was also partially funded by the Provincial Secretariat for Higher Education and Scientific Research of the Autonomous Province of Vojvodina, Republic of Serbia (Grant No.
142-451-3490/2023). We would like to thank the Tourist Organization of Zlatibor and Mr. Stojan Vuković for providing the data on tourist arrivals in Stopića cave. Author Aleksandar Antić is grateful for the postdoctoral Swiss Government
Excellence Scholarships for the academic year 2023/2024.

\bibliographystyle{plain} 

\bibliography{references}

\begin{thebibliography}{10}

\bibitem{abellana2021hybrid}
Dharyll Prince~Mariscal Abellana, Donna Marie~Canizares Rivero, Ma~Elena Aparente, and Aries Rivero.
\newblock Hybrid {SVR-SARIMA} model for tourism forecasting using {PROMETHEE II} as a selection methodology: a {P}hilippine scenario.
\newblock {\em Journal of Tourism Futures}, 7(1):78--97, 2021.

\bibitem{a11_abrate2019impact}
Graziano Abrate, Juan~Luis Nicolau, and Giampaolo Viglia.
\newblock The impact of dynamic price variability on revenue maximization.
\newblock {\em Tourism {M}anagement}, 74:224--233, 2019.

\bibitem{a26_abu2021sarima}
Noratikah Abu, Wan~Nur Syahidah, Megat~Muhammad Afif, and Syarifah~Zyurina Nordin.
\newblock {SARIMA} and {E}xponential {S}moothing model for forecasting ecotourism demand: {A} case study in {N}ational {P}ark {K}uala {T}ahan, {P}ahang.
\newblock In {\em Journal of Physics: Conference Series}, volume 1988, page 012118. IOP Publishing, 2021.

\bibitem{a24_aliani2018modeling}
Hamide Aliani, Sasan~Babaie Kafaky, Seyed~Masoud Monavari, and Kiumars Dourani.
\newblock Modeling and prediction of future ecotourism conditions applying system dynamics.
\newblock {\em Environmental monitoring and assessment}, 190:1--18, 2018.

\bibitem{athanasopoulos2009hierarchical}
George Athanasopoulos, Roman~A Ahmed, and Rob~J Hyndman.
\newblock Hierarchical forecasts for {A}ustralian domestic tourism.
\newblock {\em International Journal of Forecasting}, 25(1):146--166, 2009.

\bibitem{a57_baker1988environmental}
A.~Baker and D.~\&~Genty.
\newblock Environmental pressures on conserving cave speleothems: effects of changing surface land use and increased cave tourism.
\newblock {\em Journal of Environmental Management}, 53(2):165--175, 1988.

\bibitem{a40_bentivenga2019geoheritage}
Mario Bentivenga, Francesco Cavalcante, Giuseppe Mastronuzzi, Giuseppe Palladino, and Giacomo Prosser.
\newblock Geoheritage: {T}he foundation for sustainable geotourism, 2019.

\bibitem{box2015time}
George~EP Box, Gwilym~M Jenkins, Gregory~C Reinsel, and Greta~M Ljung.
\newblock {\em Time series analysis: forecasting and control}.
\newblock John Wiley \& Sons, 2015.

\bibitem{a45_brilha2002geoconservation}
Jos{\'e} Brilha.
\newblock Geoconservation and protected areas.
\newblock {\em Environmental conservation}, 29(3):273--276, 2002.

\bibitem{a3_buhalis2000tourism}
Dimitrios Buhalis.
\newblock Tourism and information technologies: {P}ast, present and future.
\newblock {\em Tourism recreation research}, 25(1):41--58, 2000.

\bibitem{a1_buhalis2005tourism}
Dimitrios Buhalis.
\newblock The tourism phenomenon: the new tourist and consumer.
\newblock In {\em Tourism in the {A}ge of {G}lobalisation}, pages 83--110. Routledge, 2005.

\bibitem{a47_burek2008history}
Cynthia~V Burek and Colin~D Prosser.
\newblock {\em The history of geoconservation: an introduction}, volume 300.
\newblock The Geological Society of London London, 2008.

\bibitem{a20_burger2001practitioners}
CJSC Burger, M~Dohnal, M~Kathrada, and R~Law.
\newblock A practitioners guide to time-series methods for tourism demand forecasting—a case study of {D}urban, {S}outh {A}frica.
\newblock {\em Tourism management}, 22(4):403--409, 2001.

\bibitem{a28_butler1999sustainable}
Richard~W Butler.
\newblock Sustainable tourism: {A} state-of-the-art review.
\newblock {\em Tourism geographies}, 1(1):7--25, 1999.

\bibitem{a36_carrion2021environmental}
PA{\'U}L Carri{\'o}n-Mero, Fernando Morante-Carballo, PAULA Palomeque-Ar{\'e}valo, and BORIS Apolo-Masache.
\newblock Environmental assessment and tourist carrying capacity for the development of geosites in the framework of geotourism, {G}uayaquil, {E}cuador.
\newblock {\em WIT Trans. Ecol. Environ}, 253:149--160, 2021.

\bibitem{a37_cheablam2021assessment}
Onanong Cheablam, Pavit Tansakul, Budsarin Nantakat, and Sirinan Pantaruk.
\newblock Assessment of the geotourism resource potential of the {S}atun {UNESCO} global geopark, {T}hailand.
\newblock {\em Geoheritage}, 13:1--16, 2021.

\bibitem{a42_chen2015principles}
Anze Chen, Yunting Lu, and Young~CY Ng.
\newblock {\em The principles of geotourism}.
\newblock Springer, 2015.

\bibitem{chen2012forecasting}
Chun-Fu Chen, Ming-Cheng Lai, and Ching-Chiang Yeh.
\newblock Forecasting tourism demand based on empirical mode decomposition and neural network.
\newblock {\em Knowledge-Based Systems}, 26:281--287, 2012.

\bibitem{chen2007support}
Kuan-Yu Chen and Cheng-Hua Wang.
\newblock Support vector regression with genetic algorithms in forecasting tourism demand.
\newblock {\em Tourism Management}, 28(1):215--226, 2007.

\bibitem{chen2015forecasting}
Rong Chen, Chang-Yong Liang, Wei-Chiang Hong, and Dong-Xiao Gu.
\newblock Forecasting holiday daily tourist flow based on seasonal support vector regression with adaptive genetic algorithm.
\newblock {\em Applied Soft Computing}, 26:435--443, 2015.

\bibitem{a55_chiarini2022global}
V.~Chiarini, Duckeck, J., and J.~De~Waele.
\newblock A global perspective on sustainable show cave tourism.
\newblock {\em Geoheritage}, 14(3):82, 2022.

\bibitem{clark2019bringing}
Matt Clark, Emily~J Wilkins, Dani~T Dagan, Robert Powell, Ryan~L Sharp, and Vicken Hillis.
\newblock Bringing forecasting into the future: {U}sing {G}oogle to predict visitation in {US} national parks.
\newblock {\em Journal of environmental management}, 243:88--94, 2019.

\bibitem{claveria2015tourism}
Oscar Claveria, Enric Monte, and Salvador Torra.
\newblock Tourism demand forecasting with neural network models: different ways of treating information.
\newblock {\em International Journal of Tourism Research}, 17(5):492--500, 2015.

\bibitem{a60_constantin2021monitoring}
S.~Constantin, I.~C. Mirea, A.~Petculescu, R.~A. Arghir, D.~Ș. M\u{a}ntoiu, M.~Kenesz, and O.~T. Moldovan.
\newblock Monitoring human impact in show caves. {A} study of four {R}omanian caves.
\newblock {\em Sustainability}, 13(4):1619, 2021.

\bibitem{SVR}
Corrina Cortes and Vladimir Vapnik.
\newblock Support-vector machine.
\newblock {\em Machine Learning}, 20(3):273--297, 1995.

\bibitem{a49_crofts2020guidelines}
Roger Crofts, John~E Gordon, JB~Brilha, Murray Gray, John Gunn, Jonathan Larwood, Vincent~L Santucci, Daniel Tormey, and Graeme~L Worboys.
\newblock Guidelines for geoconservation in protected and conserved areas, 2020.

\bibitem{a7_crouch1999tourism}
Geoffrey~I Crouch and JR~Brent Ritchie.
\newblock Tourism, competitiveness, and societal prosperity.
\newblock {\em Journal of business research}, 44(3):137--152, 1999.

\bibitem{a23_dimitrov2013long}
Preslav Dimitrov.
\newblock Long-run forecasting of the number of the ecotourism arrivals in the municipality of stambolovo, bulgaria.
\newblock {\em Tourism \& Management Studies}, 9(1):41--47, 2013.

\bibitem{a43_dowling2018geotourism}
Ross Dowling and David Newsome.
\newblock Geotourism: definition, characteristics and international perspectives.
\newblock {\em Handbook of geotourism}, pages 1--22, 2018.

\bibitem{a41_dowling2006geotourism}
Ross~K Dowling and David Newsome.
\newblock Geotourism's issues and challenges.
\newblock In {\em Geotourism}, pages 242--254. Routledge, 2006.

\bibitem{a5_dwyer2009destination}
Larry Dwyer, Deborah Edwards, Nina Mistilis, Carolina Roman, and Noel Scott.
\newblock Destination and enterprise management for a tourism future.
\newblock {\em Tourism Management}, 30(1):63--74, 2009.

\bibitem{a31_fennell2004tourism}
David~A Fennell and Kevin Ebert.
\newblock Tourism and the precautionary principle.
\newblock {\em Journal of Sustainable Tourism}, 12(6):461--479, 2004.

\bibitem{fildes2011evaluating}
Robert Fildes, Yingqi Wei, and Suzilah Ismail.
\newblock Evaluating the forecasting performance of econometric models of air passenger traffic flows using multiple error measures.
\newblock {\em International Journal of Forecasting}, 27(3):902--922, 2011.

\bibitem{a2_frechtling2012forecasting}
Douglas Frechtling.
\newblock {\em Forecasting tourism demand}.
\newblock Routledge, 2012.

\bibitem{a39_gordon2018geoheritage}
John~E Gordon.
\newblock Geoheritage, geotourism and the cultural landscape: {E}nhancing the visitor experience and promoting geoconservation.
\newblock {\em Geosciences}, 8(4):136, 2018.

\bibitem{a46_gray2005geodiversity}
Murray Gray.
\newblock Geodiversity and geoconservation: what, why, and how?
\newblock In {\em The George Wright Forum}, volume~22, pages 4--12. JSTOR, 2005.

\bibitem{gunter2016forecasting}
Ulrich Gunter and Irem {\"O}nder.
\newblock Forecasting city arrivals with {G}oogle {A}nalytics.
\newblock {\em Annals of Tourism Research}, 61:199--212, 2016.

\bibitem{a35_guo2017remaking}
Wei Guo and Shanshan Chung.
\newblock Remaking tourism carrying capacity frameworks for geoparks.
\newblock {\em DEStech Transactions on Social Science, Education and Human Science}, 2017.

\bibitem{harvey199310}
Andrew~C Harvey and Neil Shephard.
\newblock Structural time series models.
\newblock 1993.

\bibitem{a48_henriques2011geoconservation}
Maria~Helena Henriques, Rui~Pena dos Reis, Jos{\'e} Brilha, and Teresa Mota.
\newblock Geoconservation as an emerging geoscience.
\newblock {\em Geoheritage}, 3:117--128, 2011.

\bibitem{a13_holloway2004marketing}
J~Christopher Holloway.
\newblock {\em Marketing for tourism}.
\newblock Pearson education, 2004.

\bibitem{hu2022tourism}
Mingming Hu, Hengyun Li, Haiyan Song, Xin Li, and Rob Law.
\newblock Tourism demand forecasting using tourist-generated online review data.
\newblock {\em Tourism Management}, 90:104490, 2022.

\bibitem{a15_hudson2017marketing}
Simon Hudson.
\newblock Marketing for tourism, hospitality \& events: a global \& digital approach.
\newblock 2017.

\bibitem{ke2017lightgbm}
Guolin Ke, Qi~Meng, Thomas Finley, Taifeng Wang, Wei Chen, Weidong Ma, Qiwei Ye, and Tie-Yan Liu.
\newblock Lightgbm: {A} highly efficient gradient boosting decision tree.
\newblock {\em Advances in neural information processing systems}, 30, 2017.

\bibitem{a17_lenny2007impact}
SC~Lenny~Koh, Mehmet Demirbag, Erkan Bayraktar, Ekrem Tatoglu, and Selim Zaim.
\newblock The impact of supply chain management practices on performance of smes.
\newblock {\em Industrial management \& data systems}, 107(1):103--124, 2007.

\bibitem{li2020forecasting}
Hengyun Li, Mingming Hu, and Gang Li.
\newblock Forecasting tourism demand with multisource big data.
\newblock {\em Annals of Tourism Research}, 83:102912, 2020.

\bibitem{a10_li2019competitive}
Hui Li and Kannan Srinivasan.
\newblock Competitive dynamics in the sharing economy: {A}n analysis in the context of {A}irbnb and hotels.
\newblock {\em Marketing Science}, 38(3):365--391, 2019.

\bibitem{a30_liu2003sustainable}
Zhenhua Liu.
\newblock Sustainable tourism development: {A} critique.
\newblock {\em Journal of Sustainable Tourism}, 11(6):459--475, 2003.

\bibitem{a34_lobo2015tourist}
Heros Augusto~Santos Lobo.
\newblock Tourist carrying capacity of {S}antana cave ({PETAR-SP}, {B}razil): {A} new method based on a critical atmospheric parameter.
\newblock {\em Tourism Management Perspectives}, 16:67--75, 2015.

\bibitem{a32_lobo2013projection}
Heros Augusto~Santos Lobo, Eleonora Trajano, Maur{\'\i}cio de~Alc{\^a}ntara~Marinho, Maria~Elina Bichuette, Jos{\'e} Antonio~Basso Scaleante, Oscarlina Aparecida~Furquim Scaleante, B{\'a}rbara~Nazar{\'e} Rocha, and Francisco~Villela Laterza.
\newblock Projection of tourist scenarios onto fragility maps: {F}ramework for determination of provisional tourist carrying capacity in a {B}razilian show cave.
\newblock {\em Tourism Management}, 35:234--243, 2013.

\bibitem{a18_lovelock2013strategies}
Christopher~H Lovelock.
\newblock Strategies for managing demand in capacity-constrained service organisations.
\newblock In {\em Marketing in the Service Industries}, pages 12--30. Routledge, 2013.

\bibitem{makridakis2000m3}
Spyros Makridakis and Michele Hibon.
\newblock The {M3}-{C}ompetition: results, conclusions and implications.
\newblock {\em International Journal of Forecasting}, 16(4):451--476, 2000.

\bibitem{makridakis2020m4}
Spyros Makridakis, Evangelos Spiliotis, and Vassilios Assimakopoulos.
\newblock The {M4} {C}ompetition: 100,000 time series and 61 forecasting methods.
\newblock {\em International Journal of Forecasting}, 36(1):54--74, 2020.

\bibitem{makridakis2022m5}
Spyros Makridakis, Evangelos Spiliotis, and Vassilios Assimakopoulos.
\newblock {M5} accuracy competition: {R}esults, findings, and conclusions.
\newblock {\em International Journal of Forecasting}, 38(4):1346--1364, 2022.

\bibitem{a16_mandal2018exploring}
Santanu Mandal.
\newblock Exploring the influence of big data analytics management capabilities on sustainable tourism supply chain performance: the moderating role of technology orientation.
\newblock {\em Journal of Travel \& Tourism Marketing}, 35(8):1104--1118, 2018.

\bibitem{a9_martins2017empirical}
Lu{\'\i}s~Filipe Martins, Yi~Gan, and Alexandra Ferreira-Lopes.
\newblock An empirical analysis of the influence of macroeconomic determinants on {W}orld tourism demand.
\newblock {\em Tourism Management}, 61:248--260, 2017.

\bibitem{a29_mccool2001tourism}
Stephen~F McCool and David~W Lime.
\newblock Tourism carrying capacity: tempting fantasy or useful reality?
\newblock {\em Journal of Sustainable Tourism}, 9(5):372--388, 2001.

\bibitem{nor2018hybrid}
ME~Nor, A~IM Nurul, and MS~Rusiman.
\newblock A hybrid approach on tourism demand forecasting.
\newblock In {\em Journal of Physics: Conference Series}, volume 995, page 012034. IOP Publishing, 2018.

\bibitem{a59_novas2017real}
N.~Novas, J.~A. G\'{a}zquez, J.~MacLennan, R.~M. Garc\'{i}a, M.~Fern\'{a}ndez-Ros, and F.~Manzano-Agugliaro.
\newblock A real-time underground environment monitoring system for sustainable tourism of caves.
\newblock {\em Journal of Cleaner Production}, 142:2707--2721, 2017.

\bibitem{a44_olafsdottir2019geotourism}
Rannveig {\'O}lafsd{\'o}ttir.
\newblock Geotourism, 2019.

\bibitem{a27_o1986tourism}
Ainsley~M O'Reilly.
\newblock Tourism carrying capacity: {C}oncept and issues.
\newblock {\em Tourism Management}, 7(4):254--258, 1986.

\bibitem{park2017short}
Sangkon Park, Jungmin Lee, and Wonho Song.
\newblock Short-term forecasting of {J}apanese tourist inflow to {S}outh {K}orea using {G}oogle trends data.
\newblock {\em Journal of Travel \& Tourism Marketing}, 34(3):357--368, 2017.

\bibitem{a56_pulido1997human}
Pulido-Bosch, A., Martin-Rosales, W., L\'{o}pez-Chicano, M., Rodriguez-Navarro, C.~M., and A.~Vallejos.
\newblock Human impact in a tourist karstic cave ({A}racena, {S}pain).
\newblock {\em Environmental geology}, 31(3):142--149, 1997.

\bibitem{a25_rice2019forecasting}
William~L Rice, So~Young Park, Bing Pan, and Peter Newman.
\newblock Forecasting campground demand in us national parks.
\newblock {\em Annals of Tourism Research}, 75:424--438, 2019.

\bibitem{a4_ritchie2003competitive}
JR~Brent Ritchie and Geoffrey~Ian Crouch.
\newblock {\em The competitive destination: {A} sustainable tourism perspective}.
\newblock Cabi, 2003.

\bibitem{a51_ruban2018karst}
D.~A. Ruban.
\newblock Karst as important resource for geopark-based tourism: {C}urrent state and biases.
\newblock {\em Resources}, 7(4):82, 2018.

\bibitem{a19_shabanpour2018analysis}
Ramin Shabanpour, Nima Golshani, Mohammad Tayarani, Joshua Auld, and Abolfazl~Kouros Mohammadian.
\newblock Analysis of telecommuting behavior and impacts on travel demand and the environment.
\newblock {\em Transportation Research Part D: Transport and Environment}, 62:563--576, 2018.

\bibitem{a14_sigala2012social}
Marianna Sigala, Evangelos Christou, and Ulrike Gretzel.
\newblock {\em Social media in travel, tourism and hospitality: {T}heory, practice and cases}.
\newblock Ashgate Publishing, Ltd., 2012.

\bibitem{song2019review}
Haiyan Song, Richard~TR Qiu, and Jinah Park.
\newblock A review of research on tourism demand forecasting: {L}aunching the {A}nnals of {T}ourism {R}esearch {C}urated {C}ollection on tourism demand forecasting.
\newblock {\em Annals of Tourism Research}, 75:338--362, 2019.

\bibitem{a8_song2006tourism}
Haiyan Song, Lindsay Turner, et~al.
\newblock Tourism demand forecasting.
\newblock {\em International handbook on the economics of tourism}, pages 89--114, 2006.

\bibitem{a12_song2012tourism}
Haiyan Song and Stephen~F Witt.
\newblock {\em Tourism demand modelling and forecasting}.
\newblock Routledge, 2012.

\bibitem{sun2019forecasting}
Shaolong Sun, Yunjie Wei, Kwok-Leung Tsui, and Shouyang Wang.
\newblock Forecasting tourist arrivals with machine learning and internet search index.
\newblock {\em Tourism Management}, 70:1--10, 2019.

\bibitem{a38_sunkar2022geotourism}
Arzyana Sunkar, Anindika~Putri Lakspriyanti, Eko Haryono, Mohsen Brahmi, Pindi Setiawan, and Aziz~Fardhani Jaya.
\newblock Geotourism hazards and carrying capacity in geosites of {S}angkulirang-{M}angkalihat {K}arst, {I}ndonesia.
\newblock {\em Sustainability}, 14(3):1704, 2022.

\bibitem{a54_taheri2021human}
C.~Taheri, K.and~Groves.
\newblock Human-{K}arst {L}andscape {I}nteractions and the {A}nthropo-{K}arstosphere: {T}oward a {N}exus of {G}eoethics, {G}roundwater, and a {S}ustainable {S}ociety.
\newblock pages 231--236, 2021.

\bibitem{taylor2018forecasting}
Sean~J Taylor and Benjamin Letham.
\newblock Forecasting at scale.
\newblock {\em The American Statistician}, 72(1):37--45, 2018.

\bibitem{a52_telbisz2020significance}
T.~Telbisz and L~Mari.
\newblock The significance of karst areas in {E}uropean national parks and geoparks.
\newblock {\em Open Geosciences}, 12(1):117--135, 2020.

\bibitem{a21_trauer2006conceptualizing}
Birgit Trauer.
\newblock Conceptualizing special interest tourism—frameworks for analysis.
\newblock {\em Tourism Management}, 27(2):183--200, 2006.

\bibitem{triebe2021neuralprophet}
Oskar Triebe, Hansika Hewamalage, Polina Pilyugina, Nikolay Laptev, Christoph Bergmeir, and Ram Rajagopal.
\newblock Neuralprophet: {E}xplainable forecasting at scale.
\newblock {\em arXiv preprint arXiv:2111.15397}, 2021.

\bibitem{triebe2019ar}
Oskar Triebe, Nikolay Laptev, and Ram Rajagopal.
\newblock Ar-net: {A} simple auto-regressive neural network for time-series.
\newblock {\em arXiv preprint arXiv:1911.12436}, 2019.

\bibitem{a6_vanhove2022economics}
Norbert Vanhove.
\newblock {\em The economics of tourism destinations: {T}heory and practice}.
\newblock Routledge, 2022.

\bibitem{volchek2019forecasting}
Katerina Volchek, Anyu Liu, Haiyan Song, and Dimitrios Buhalis.
\newblock Forecasting tourist arrivals at attractions: {S}earch engine empowered methodologies.
\newblock {\em Tourism Economics}, 25(3):425--447, 2019.

\bibitem{a58_sebela2015cave}
S.~\v{S}ebela, J.~Turk, and T~Pipan.
\newblock Cave micro-climate and tourism: towards 200 years (1819–2015) at {P}ostojnska jama ({S}lovenia).
\newblock {\em Cave and Karst Science}, 42(2):75--85, 2015.

\bibitem{a50_williams2020geoconservation}
M.~A. Williams, M.~T. McHenry, and A.~Boothroyd.
\newblock Geoconservation and geotourism: {C}hallenges and unifying themes.
\newblock {\em Geoheritage}, 12(3):63, 2020.

\bibitem{a22_xie2021forecasting}
Gang Xie, Yatong Qian, and Shouyang Wang.
\newblock Forecasting chinese cruise tourism demand with big data: {A}n optimized machine learning approach.
\newblock {\em Tourism Management}, 82:104208, 2021.

\bibitem{a33_zelenka2014concept}
Josef Zelenka and Jaroslav Kacetl.
\newblock The concept of carrying capacity in tourism.
\newblock {\em Amfiteatru Economic Journal}, 16(36):641--654, 2014.

\bibitem{a53_zhang2023aesthetic}
S.~Zhang, K.~Xiong, G.~Fei, H.~Zhang, and Y.~Chen.
\newblock Aesthetic value protection and tourism development of the world natural heritage sites: a literature review and implications for the world heritage karst sites.
\newblock {\em Heritage Science}, 11(1):30, 2023.

\end{thebibliography}

\end{document}